\PassOptionsToClass{10pt}{revtex4-2}
\documentclass[aps,prl,showpacs,floatfix,twocolumn,byrevtex,superscriptaddress]{revtex4-1}

\usepackage{amsmath}
\usepackage{amssymb}
\usepackage{amstext}
\usepackage{amsopn}
\usepackage{amsfonts}
\usepackage{amsxtra}
\usepackage[english]{babel}
\usepackage{graphicx}
\usepackage{bm}
\usepackage{multirow}
\usepackage{dcolumn}
\usepackage{color}
\usepackage{hyperref}
\usepackage{todonotes}
\usepackage{verbatim}
\usepackage{soul}

\def\mathbi#1{\ensuremath{\textbf{\em #1}}}
\def\MBT{MnBi$_2$Te$_4$}

\begin{document}

\title{Coexistence of Surface Ferromagnetism and Gapless Topological State in \MBT{}} 

\author{D. Nevola}\email[]{nevola@bnl.gov}
\affiliation{Condensed Matter Physics and Materials Science Department, Brookhaven National Laboratory, Upton, New York 11973, USA}
\author{H. X. Li}
\affiliation{Materials Science and Technology Division, Oak Ridge National Laboratory, Oak Ridge, Tennessee 37831, USA}
\author{J.-Q. Yan}
\affiliation{Materials Science and Technology Division, Oak Ridge National Laboratory, Oak Ridge, Tennessee 37831, USA}
\author{R. G. Moore}
\affiliation{Materials Science and Technology Division, Oak Ridge National Laboratory, Oak Ridge, Tennessee 37831, USA}
\author{H.-N. Lee}
\affiliation{Materials Science and Technology Division, Oak Ridge National Laboratory, Oak Ridge, Tennessee 37831, USA}
\author{H. Miao}\email[]{miaoh@ornl.gov}
\affiliation{Materials Science and Technology Division, Oak Ridge National Laboratory, Oak Ridge, Tennessee 37831, USA}
\author{P. D. Johnson}\email[]{pdj@bnl.gov}
\affiliation{Condensed Matter Physics and Materials Science Department, Brookhaven National Laboratory, Upton, New York 11973, USA}

\date{\today}


\date{\today}

\begin{abstract}
Surface magnetism and its correlation with the electronic structure are critical to understand the gapless topological surface state in the intrinsic magnetic topological insulator \MBT{}. Here, using static and time resolved angle-resolved photoemission spectroscopy (ARPES), we find a significant ARPES intensity change together with a gap opening on a Rashba-like conduction band. Comparison with a model simulation strongly indicates that the surface magnetism on cleaved \MBT{} is the same as its bulk state. The coexistence of surface ferromagnetism and a gapless TSS uncovers the novel complexity of \MBT{} that may be responsible for the low quantum anomalous Hall temperature of exfoliated \MBT{}.

\end{abstract}

\maketitle
The combination of magnetism and nontrivial topology opens new routes to realize novel quantum phenomena such as the quantum anomalous Hall effect (QAHE) and axion electrodynamics \cite{Li2010, Yu2010, Chang2013, Checkelsky2014, Wang2015, Morimoto2015, Mogi2017, Tokura2019, Li2019, Xu2019, Zhang2019, Nie2020}. The discovery of quantized Hall resistivity in Cr$_{0.15}$(Bi$_{0.1}$Sb$_{0.9}$)$_{1.85}$Te$_{3}$ demonstrates the first realization of the QAHE without landau levels \cite{Chang2013}. However, the temperature at which the QAHE has been observed is below 100~mK, two orders of magnitude lower than the ferromagnetic ordering temperature, which could be due to the large real space electronic inhomogeneity \cite{Lee2015, Tokura2019}. The desire to achieve a higher QAHE temperature has thus motivated the search for intrinsic magnetic topological insulators (TIs). 

\begin{figure}[tb]
\includegraphics[width=8.6 cm]{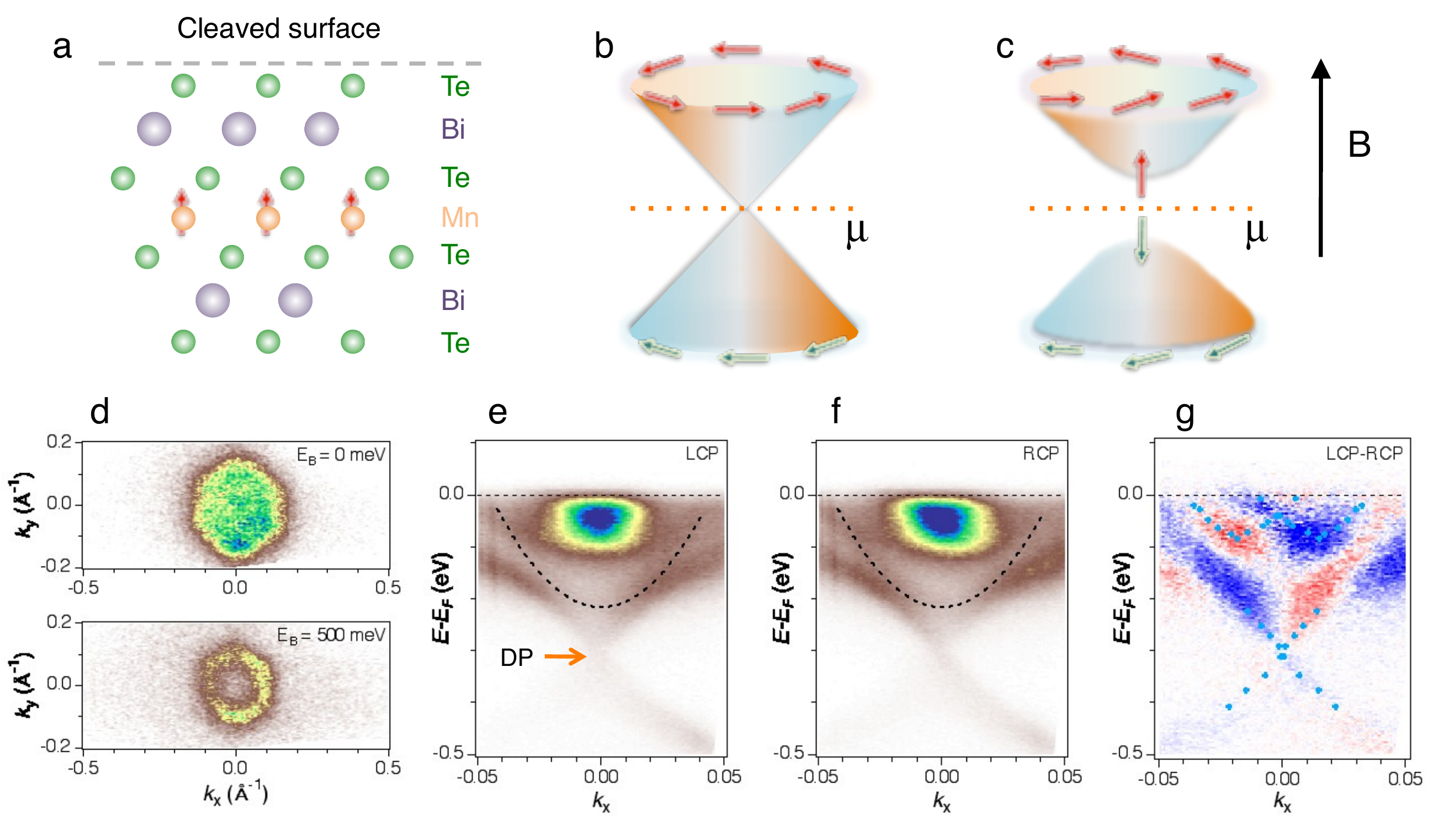}
\caption{Crystal and band structure of the intrinsic magnetic TI \MBT{}. (a) The SL of \MBT{}. Mn carries large magnetic moment that orders ferromagnetically within the same Mn atomic-layer and antiferromagnetically between two adjacent Mn atomic-layers. Schematic plots of the topological surface state without (b) and with (c) effective ferromagnetic field. (d) ARPES constant energy plot taken with He-I$\alpha=$21.2 eV at $E_{F}$ and $E_{B}$=0.5~eV. (e) and (f) are ARPES intensity plots measured with LCP and RCP across the $\Gamma$ point. The red arrow in (e) indicates the surface Dirac point at $E_{B}$=0.3~eV. The black curve acts as a guide to the eye and indicates the position of the bulk conduction band, which has $p_z$ orbital character. The intensity difference between LCP and RCP is shown in (g). The cyan markers indicate the Rashba-like conduction band and massless Dirac cone. The ARPES spectra is taken in the Paramagnetic state at 40~K.}
\label{Fig1}
\end{figure}

The A-type antiferromagnetic \MBT{} was recently proposed as the first intrinsic magnetic TI \cite{Li2019,Zhang2019, Chen2019, Chen2019b, Yan2019,LiH2019, Hao2019, Rienks2019, Otrokov2019}. Its crystal structure is shown in Fig.~\ref{Fig1}a, where each septuple layer (SL) consists of alternating Bi and Te layers, similar to that in Bi$_{2}$Te$_{3}$, but with an additional Mn-Te bilayer in the middle. Magnetism originates from the Mn atoms that ferromagnetically order in the ab-plane and antiferromagnetically order along the (0001) direction \cite{Yan2019, Yan2019b}. Similar to the non-magnetic Bi$_{2}$Te$_{3}$, the topological surface state (TSS) arises from the band inversion between the Bi and Te $p_z$ bands at the $\Gamma$ point. The magnetic order induced below $T_N\sim$25~K is expected to open a gap in the massless Dirac-cone on the (0001) surface (Fig.~\ref{Fig1}b and c). This gap opening is thought to induce QAH or axion states, which form depending on the number of SLs \cite{Zhang2019, Chen2019b, Xu2019}. More recently, the quantized Hall resistivity has been observed in exfoliated \MBT{} \cite{Liu2020, Deng2020}. However, angle-resolved photoemission spectroscopy (ARPES) on the tellurium-terminated (0001) surface (Fig.~\ref{Fig1}a) shows that the gapless surface state is robust to magnetic ordering \cite{LiH2019, Chen2019, Li2019, Chen2019b, Hao2019, Swatek2019}, inconsistent with first principle calculations and transport measurements \cite{Zhang2019,Li2019, Chen2019b, Deng2020, Liu2020}. This contradiction raises the question of whether or not the surface is magnetically ordered below the bulk $T_N$ \cite{LiH2019, Chen2019, Li2019, Chen2019b, Hao2019, Swatek2019}. Here, we use static and time resolved ARPES (trARPES) with full polarization control to prove that the magnetic order is retained at the surface and fails to induce a gap in the TSS. Our results uncover a novel complexity of the intrinsic magnetic TI that may be responsible for the low QAH temperature in \MBT{}.

%

The laser based ARPES was performed at Brookhaven National Laboratory using an SES-2002 analyzer. The He-I$\alpha$ measurement was performed at Oak Ridge National Laboratory using a DA30 analyzer. The samples were cleaved in-situ and experiments were performed at a base pressure of 4$\times$10$^{-11}$ torr. For the laser based source, we used the fourth harmonic of a 800~nm beam operating at 250~kHz. The output of a Ti:Sapphire oscillator (Coherent Inc. Vitara T) was used to seed a regenerative amplifier (Coherent Inc. RegA 9050) in order to produce pulses of 30~nm bandwidth and 70~fs. Fourth harmonic generation was then achieved through a series of nonlinear processes using $\beta$-BaB$_{2}$O$_{4}$ crystals. The fourth harmonic was subsequently compressed using a prism compressor. For the trARPES measurements, the 800~nm fundamental pulses were used to pump the sample and a delay stage was used to control the delay time in between the two pulses. The Fermi level location was calibrated using photoemission from polycrystalline gold. The pump-probe temporal overlap and resolution was determined through measurements at high energies in Bi$_{2}$Se$_{3}$. For the laser based ARPES measurements, the resolution was 29~meV, and for the trARPES measurements, the energy and time resolutions were 50~meV and 180~fs, respectively.

The band structure of \MBT{} consists of several bands within the first few hundred meV of $E_{F}$. Our ARPES spectra (Fig.~\ref{Fig1}d-f) agree with previous measurements, placing the center of the TSS, the Dirac point (DP), approximately 300~meV below $E_{F}$ with a dispersive bulk conduction band 80~meV above the DP. In addition, a feature with large intensity within 100~meV of $E_{F}$ has been observed at the zone center \cite{Hao2019, Chen2019}. This feature is also clear in the constant energy maps taken with He-I$\alpha$, where we observe a hexagonal surface without a density of states at the zone center 200~meV below the DP, and a hexagonal surface with a density of states at the zone center at $E_{F}$ (Fig.~\ref{Fig1}d). Since the intensity of this feature is known to be enhanced at low photon energies \cite{Chen2019, Hao2019}, we use laser photoemission to probe its properties and temperature dependence. 

%
\begin{figure}[tb]
\includegraphics[width=8.6 cm]{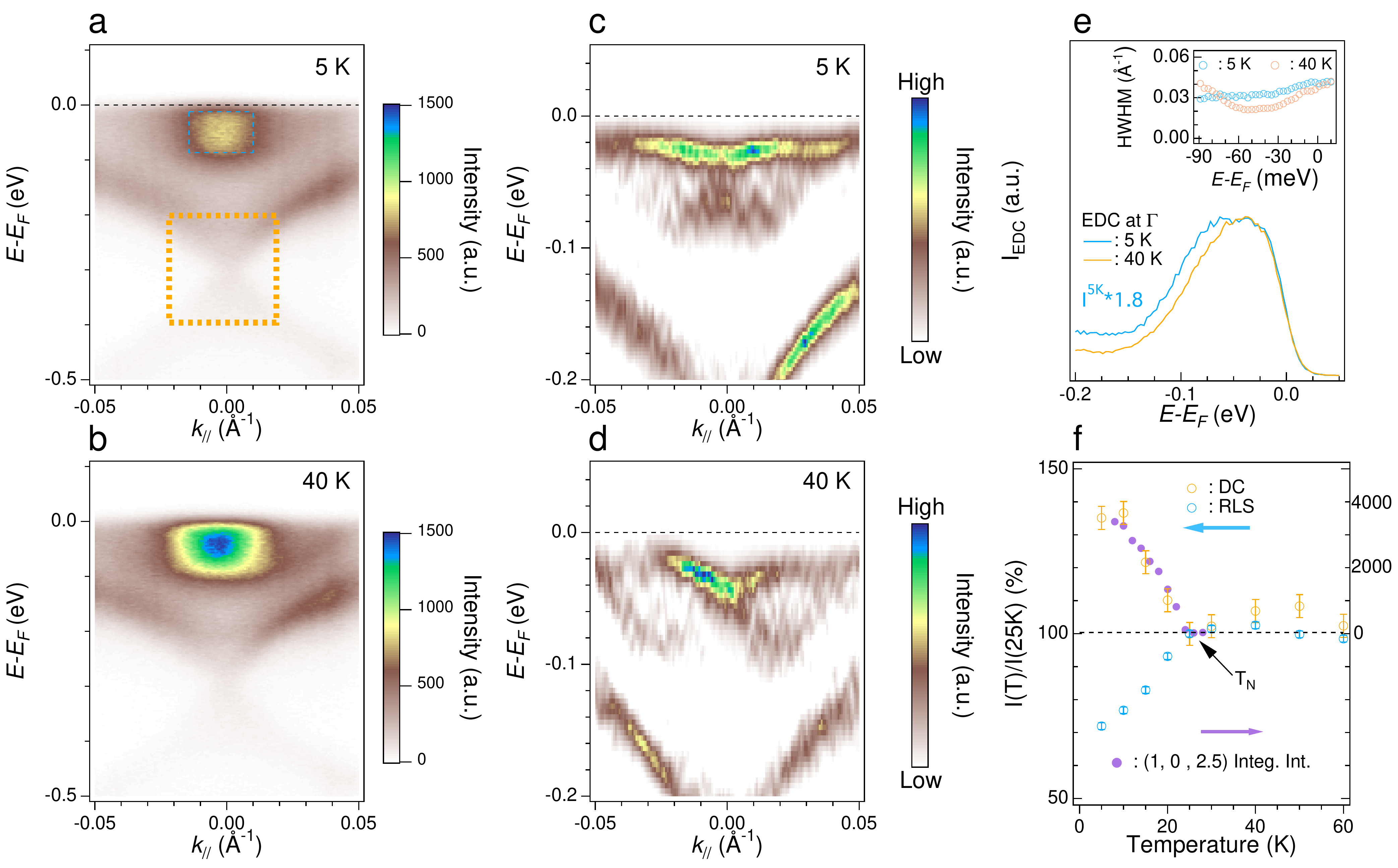}
\caption{Evidence of surface magnetic order below $T_N$. (a) and (b) are ARPES intensity plot at 5 and 40~K, respectively. These plots are obtained by adding LCP and RCP to reduce the polarization matrix elements effects. (c) and (d) are the curvature plots of (a) and (b) in a narrower energy range. Cyan curves act as guide to the eye indicate the gapped Rashba-like band at 5~K. Due to the presence of the bulk state, the Rashba-like splitting derived from the curvature analysis is slightly larger than that shown in Fig.~\ref{Fig1}g. A direct comparison of EDCs below (5~K) and above (40~K) $T_N$ at the $\Gamma$ point is shown in (e). The inset compares the extracted MDC widths of the Rashba-like band at 5~K and 40~K. (f) shows the temperature dependent intensity change of the Rashba-like band (cyan) and the TSS (orange). The integration areas are shown in (a). Purple circles are the integrated intensity of magnetic Bragg peak at $\mathbi{Q}$=(1, 0, 2.5) \cite{Yan2019b}.}
\label{Fig2}
\end{figure}

We first examine the circular dichroism (CD) of these bands above $T_{N}$. Figure~\ref{Fig1}g shows the band structure difference between the left (LCP) and right (RCP) circularly polarized light, where we observe a rich chirality dependence. Such a chirality dependence observed by CD is known to be a indirect evidence of a spin separation \cite{Wang2013}. Here we focus on the Dirac-cone and the high intensity feature near $E_{F}$. First, each half of the Dirac cone shows the opposite chiral dependence, expected for a topological surface state. This result agrees with the direct spin-dependent measurement using spin-ARPES \cite{Vidal2019}. Second and most importantly, we find that the high intensity feature consists of two bands that are separated in $\mathbf{k}$, reminiscent of a Rashba state \cite{Wang2013}, where the twofold spin degeneracy is lifted due to the breaking of inversion symmetry at the surface. Since the bulk crystal structure of \MBT{} preserves the inversion symmetry, such a Rashba-like state is absent in any first-principle calculations of the bulk electronic structure. We thus conclude the Rashba-like state has a surface origin. Indeed, the clear CD is consistent with spin-separation reported in Rashba systems and TSSs.\cite{Wang2013,Rameau2019, Zhang2018}.

%
\begin{figure*}[tb]
\includegraphics[width=17.2 cm]{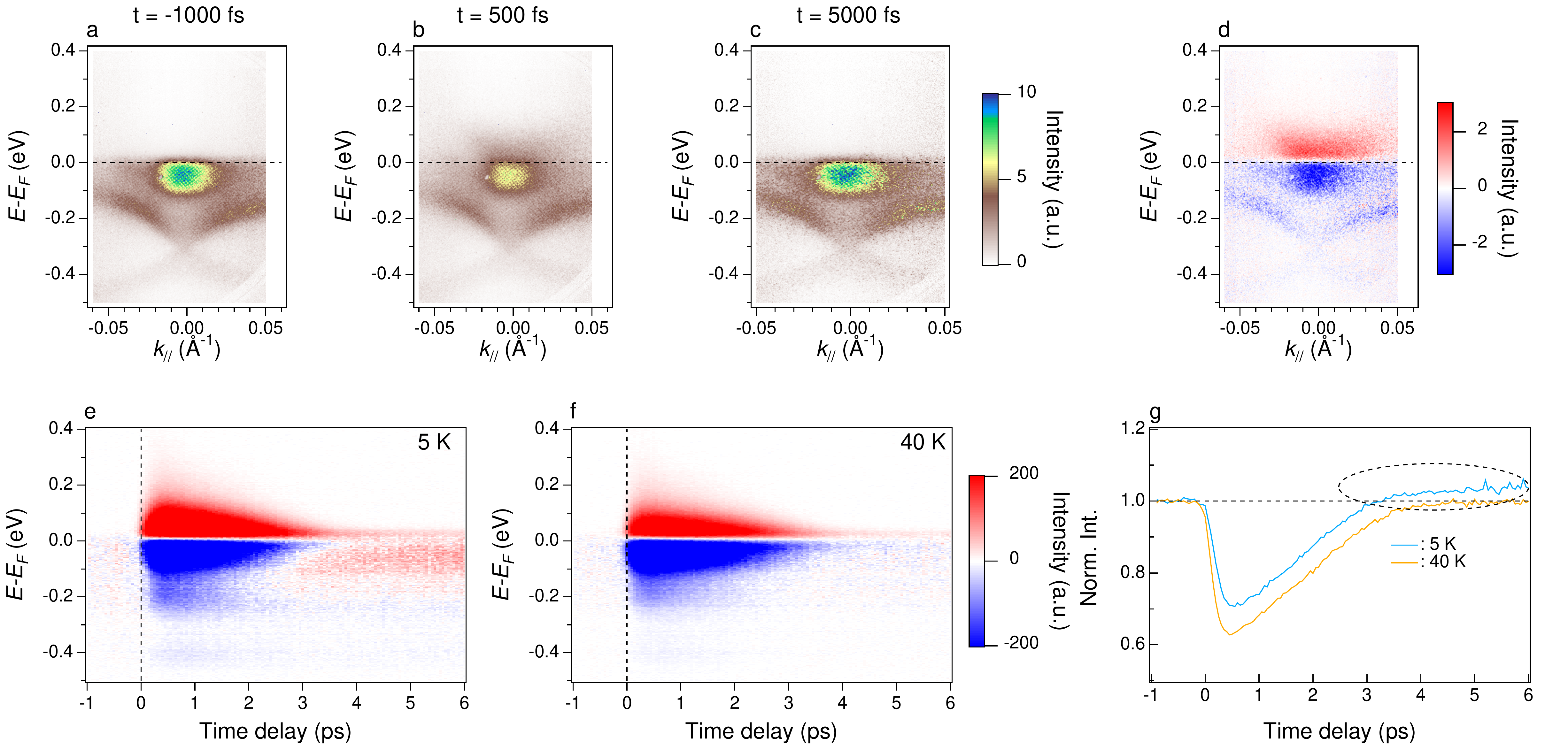}
\caption{Intensity anomaly revealed by trARPES. (a) - (c) trARPES intensity plot with different time delay. The spectra were collected at 5~K. (d) shows the representative intensity difference between (a) and (b). (e) and (f) are the $k$-integrated spectra as a function of delay time taken at 5~K and 40~K respectively. (g) Integrated intensity of the Rashba-like as function of time delay. Cyan and yellow curves represent 5~K and 40~K respectively. The dashed oval highlights the intensity increase after 4~ps. }
\label{Fig3}
\end{figure*}

We then turn to explore the temperature dependence of the Rashba-like state and TSS. Figures~\ref{Fig2}a and b show the raw data acquired as the sum of LCP and RCP below and above the magnetic ordering temperature, respectively. Before moving to discuss the Rashba-like state, we note that we do not observe a gap opening at the DP deep below the bulk $T_N$, in agreement with previous reports\cite{Chen2019,Swatek2019,Hao2019}. The Rashba-like state, however, shows some subtle differences. Upon taking the curvature analysis above the magnetic ordering temperature (Fig.~\ref{Fig2}d) \cite{Zhang2011}, we confirm the Rashba-like splitting shown in Fig.~\ref{Fig1}g. The observed momentum and energy splittings are approximately 0.025 \AA$^{-1}$ and 50~meV respectively. Taking the curvature analysis below magnetic ordering (Fig.~\ref{Fig2}c) shows a band gap of about 35~meV at the Kramers point, in agreement with a recent report \cite{Estyunin2020}. More concrete evidence of the changes to the Rashba-like bands with magnetic ordering can be given directly from the raw data. Taking an energy distribution curve (EDC) at the $\Gamma$ point above and below the magnetic ordering (Fig.~\ref{Fig2}e), we immediately see a qualitative difference in that the band becomes broader at the low temperature, opposite to what is expected from a typical temperature broadening. Additionally the EDC peak maximum flattens, offering further supporting evidence that the Kramers point splits into two bands. As a third method of analysis, we use the Lucy-Richardson deconvolution algorithm, which again supports the opening of a gap \cite{Rameau2010} (see supplemental materials). Similarly, at 40~K the extracted half-width-at-half-maximum (HWHM) of the momentum distribution curve (MDC) at the $\Gamma$ point reaches a minimum around 50~meV consistent with the band crossing shown in Fig.~\ref{Fig2}d. However at 5~K, the extracted HWHM remains flat over the entire energy range, suggesting a different behavior. This behavior is consistent with an avoided band crossing combined with a lifetime significantly broader than the gap size (see supplemental materials).

In addition to the gap opening, we observe significant intensity changes to both the Rashba-like state and TSS below and above $T_{N}$, as is immediately noticeable in the intensity plots (Fig.~\ref{Fig2}a, b). To quantify this effect, we plot the integrated intensities in the dashed rectangular areas shown in Fig.~\ref{Fig2}b as a function of temperature (Fig.~\ref{Fig2}f). We observe a clear temperature dependence and see that magnetic order has the opposite effects on the Dirac cone and the Rashba-like state,$\textit{i.e.}$ magnetic order suppresses the intensity in the Rashba-like state and enhances that in the Dirac cone. These intensity change agree with the temperature-dependent bulk magnetic order-parameter determined by neutron scattering \cite{Yan2019}. Since both of these states are of surface origin, Fig.~\ref{Fig2}f provides strong evidence that the surface electronic states ``feel'' the bulk magnetic ordering. 

To further demonstrate the spectral weight change, we perform trARPES study both above and below $T_N$. A summary of our findings is shown in Fig.~\ref{Fig3}e-g, where we display the k-integrated spectra as a function of delay time. In Figs.~\ref{Fig3}a-c, we show the ARPES intensity plots at different time delays. Fig.~\ref{Fig3}d shows a representative intensity difference between 500~fs and -1000~fs. Red and blue color represent population (intensity increase) and depopulation (intensity decrease) respectively with respect to the spectrum prior to the introduction of the pump pulse. When the system is pumped at an initial temperature of 40~K (Fig.~\ref{Fig3}f), we clearly see that the system relaxes back to its initial value after a delay time of approximately 4~ps, albeit at a slightly higher temperature (the higher temperature can be seen by noting the slight positive signal just above and the slight negative signal just below $E_F$). This can be understood qualitatively by the two-temperature model, whereby the pump pulse induces a rapid increase in the electronic temperature that ultimately equilibrates with the lattice temperature through electron-phonon scattering, resulting in a steady state equal to that prior to the pump, but at a slightly higher temperature (this equilibrium is reached after $\sim4~$ps according to our data) \cite{Allen1987,Rettig2013}. In Fig.~\ref{Fig3}e, we show the same data as Fig.~\ref{Fig3}f, but pumped while the system is magnetically ordered. At long delays above 4~ps, we observe the same temperature increase that is seen in the data above $T_N$ indicated by the slight intensity increase above $E_F$. More interestingly, we find an intensity increase within 150~meV of $E_F$, corresponding to the location of the Rashba-like state. In order to track this more clearly, we integrate over the Rashba-like state and display the resulting intensity as a function of delay time (Fig.~\ref{Fig3}g). We can see that the intensity of the band increases by approximately 3\% compared to its negative time value. This observation is qualitatively similar to Fig.~\ref{Fig2}f, where we showed the equilibrium changes to the spectral weight as a function of temperature. The 3\% change in the Rashba-like state signify a temperature increase of only a few degrees when compared with Fig.~\ref{Fig2}f. Thus, two different experiments show the same spectral weight change in the ARPES spectra with magnetic ordering.

%
\begin{figure}[tb]
\includegraphics[width=8.6 cm]{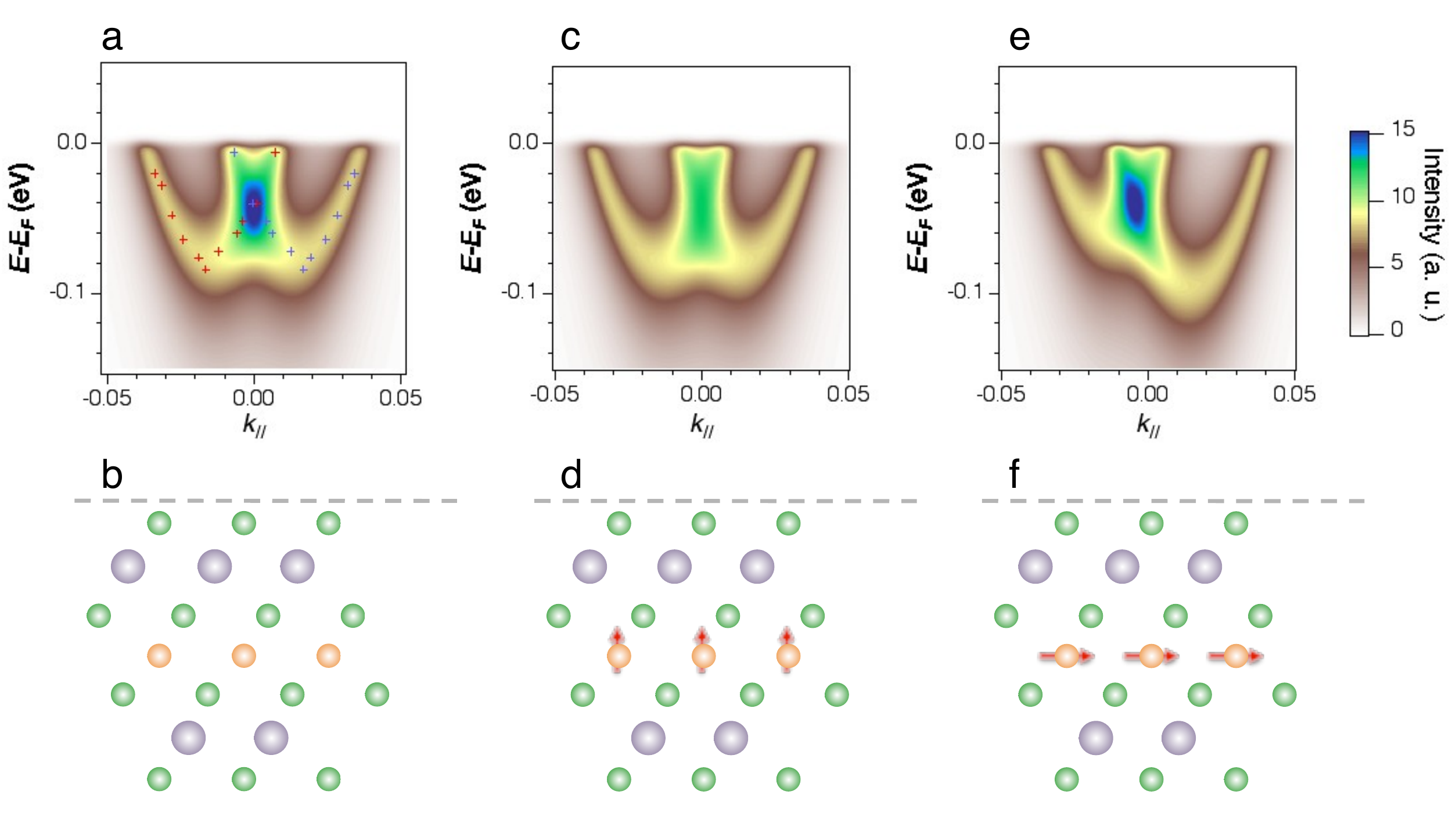}
\caption{Surface magnetism of \MBT{}. (a) Simulated spectral function of non-magnetic or antiferromagnetic surface. The non-magnetic case is shown in (b). Simulated spectral function of FM surface (d) is shown in (c). (e) and (f) FM surface with in-plane spin orientation.}
\label{Fig4}
\end{figure}

Both the ARPES intensity change and the gap opening on the Rashba-like state strongly indicate that the surface of cleaved \MBT{} is magnetically ordered. This conclusion is consistent with magnetic susceptibility measurements of exfoliated \MBT{} \cite{Liu2020, Deng2020, Chen2019b} and a magnetic force microscopy study of cleaved bulk \MBT{} \cite{Sass2020}. Recently, it has been speculated that the top SL may favor a different type of magnetic ordering, such as the G-type antiferromagnetic order \cite{Swatek2019} or FM order with spin orientation different from that in the bulk \cite{Hao2019}. To check these possibilities, we simulate our 40~K data with a simple Rashba Hamiltonian using our extracted parameters and explore the effects of different magnetic ordering. At 40~K, the effective Hamiltonian can be written as:

\begin{equation}
\begin{aligned}
H_{R}(\textbf{k})&=(\frac{\hbar^{2}k^{2}}{2m^{*}}-\mu)\mathbf{I}_{2\times2}+v_{F}(k_{y}\sigma_{x}-k_{x}\sigma_{y}) \\
&=\begin{pmatrix} 
\frac{\hbar^{2}k^{2}}{2m^{*}}-\mu & v_{F}(k_{x}+ik_{y}) \\ v_{F}(k_{x}-ik_{y}) & \frac{\hbar^{2}k^{2}}{2m^{*}}-\mu 
\end{pmatrix}
\label{Rashba}
\end{aligned}
\end{equation}
Where $\mathbf{I}_{2\times2}$ and $\sigma_{i}$ (i=x,y,z) are 2-by-2 unit matrix and the Pauli matrices respectively. The eigenvalues of Eq.~\ref{Rashba} are $E(\mathbf{k})_{\pm}=\frac{\hbar^{2}k^{2}}{2m^{*}}-\mu\pm v_{F}|\mathbf{k}|$. Figure~\ref{Fig4} shows the simulated single-particle spectral function:

\begin{equation}
A(\mathbf{k},\omega)=\frac{1}{\pi}\frac{\Gamma}{((\omega-E(\mathbf{k}))^{2}+\Gamma^{2})}
\label{Akw}
\end{equation}
where $\Gamma$=0.04 eV is the quasiparticle lifetime. $E(\mathbf{k})$ is determined by fitting of the extracted Rashba-like band dispersion. In the G-type AFM phase, the surface unit cell is doubled, however the Rashba band remains gapless \cite{Hao2019, Swatek2019}. Therefore, the observation of the gapped Rashba-like band below $T_{N}$ excludes the surface AFM order. For the FM order, the effective Hamiltonian can be written as:

\begin{equation}
H_{FM}(\textbf{k})=H_{R}(\textbf{k})+\sum_{i}\Delta_{i}\sigma_{i}
\label{RashbaM}
\end{equation}
The eigenvalues of Eq.~\ref{RashbaM} are $E(\mathbf{k})_{\pm}=\frac{\hbar^{2}k^{2}}{2m^{*}}-\mu\pm\sqrt{\Delta_{z}^2+(v_{F}k_{x}+\Delta_{x})^2+(v_{F}k_{y}-\Delta_{y})^2}$. Figure~\ref{Fig4}c and e show the simulated $A(\mathbf{k},\omega)$ with $\mathbf{\Delta}=(0,0,\Delta_{z})$ and $(\Delta_{x},0,0)$, respectively. It is clear that the out-of-plane spin orientation best describes our data, while the in-plane spin orientation shifts the two Rashba band in opposite energy directions, inconsistent with our observations. Interestingly, the simulation also shows about a 30\% intensity drop of the Rashba band that is again consistent with experimental observations. 

Our observations have important implications for the topological states in \MBT{}. Unlike non-magnetic TIs, in the magnetically ordered \MBT{}, the topological invariance along the (0001) direction is protected by the combined operator $S=\Theta\tau_{1/2}$, where $\Theta$ and $\tau_{1/2}$ are the time-reversal operator, and the translation operator along the (0001) direction by half unit cell. Since the $\tau_{1/2}$ is explicitly broken on the Te terminated (0001) surface, our observation of the surface FM with z-direction spin orientation requires additional symmetry to protect the gapless TSS. Another implication of our study is that the observation of the gapped TSS in exfoliated \MBT{} may simply arises from the coupling between the top and bottom layers \cite{Yu2010}, where the magnetic ordering effect might be less profound. Indeed, the QAH temperature of the exfoliated \MBT{} is about $\sim$1.5~K, still one order of magnitude lower than the bulk $T_{N}$ \cite{Deng2020}. A full understanding of the coexistence of surface FM and gapless TSS may thus provide a new pathway to further increase the QAH temperature in \MBT{}. 

In summary, by analyzing the temperature and time-dependent ARPES spectra, we find a strong FM effect on surface-related electronic states. Our results uncover a novel complexity of the intrinsic magnetic TI that may responsible for the yet low QAH temperature in \MBT{}.

H. M. acknowledges M. H. Du, H. X. Fu, S. Okamoto, B. H. Yan and T. T. Zhang for insightful discussions. Work at Brookhaven National Laboratory was supported by the U.S. Department of Energy, Office of Science, Office of Basic Energy Sciences, under Contract No. DE-SC0012704. Work at Oak Ridge National Laboratory (ORNL) was supported by the U.S. Department of Energy, Office of Science, Materials Sciences and Engineering Division. H. M. is supported by the Laboratory Directed Research and Development (LDRD) of ORNL, under project No.~10018. R. G. M. is supported by the LDRD of ORNL, under project No.~9983.  


\bibliography{ref}

\end{document}